\documentstyle[12pt]{article}
\newcommand{\be}{\begin{equation}}
\newcommand{\ee}{\end{equation}}
\newcommand{\bea}{\begin{eqnarray}}
\newcommand{\eea}{\end{eqnarray}}
\newcommand{\ba}{\begin{array}}
\newcommand{\ea}{\end{array}}
\newcommand{\bmat}{\left(\ba}
\newcommand{\emat}{\ea\right)}

\newcommand{\norsl}{\normalsize\sl}
\newcommand{\norsc}{\normalsize\sc}

\begin{document}



\title{Analysis of the spin structure function $g_2(x,Q^2)$\\
 and twist-3 operators
\thanks{Talk presented at {\sl YITP Workshop on
``From Hadronic Matter to Quark Matter: Evolving View of Hadronic Matter"}
YITP, Kyoto Japan,October 1994}
}

\author{
\norsc    Yoshiaki YASUI\\
\norsl  Dept. of Physics, Hiroshima University\\
\norsl  Higashi-Hiroshima 724, JAPAN}

\date{}

\maketitle

\begin{abstract}
We discuss the spin-dependent structure function $g_2(x,Q^2)$
in the framework of the operator product expansion.
It is noted that the anomalous dimensions and coefficient
functions for the twist-3 gluon-field-dependent operators
depend on the choice of the operator basis.
The role played by the operators proportional to the equation of
motion is clarified.

\end{abstract}

\begin{picture}(5,2)(-340,-375)
\put(2.3,-80){HUPD-9503}
\put(2.3,-95){January, 1995}
\end{picture}

%
\baselineskip 24pt

\section{Introduction}

The spin structure of nucleon is described by the two
spin-dependent structure functions $g_1(x,Q^2)$ and $g_2(x,Q^2)$.
Recent experiments  at CERN \cite {EMC,SMC}
and SLAC \cite{E142} have provided new data of $g_1(x,Q^2)$
spin structure function and
these data have prompted many authors to reanalyze
 $g_1(x,Q^2)$ in the connection with the Bjorken sum rule~\cite{UEMA}.
The measurement of $g_2(x,Q^2)$ was also proposed  at CERN
and SLAC. The first experimental data was published by the SMC group at
CERN \cite{G2N}.
Now theoretical investigations on $g_2(x,Q^2)$ structure
function become more and more important.

For $g_1(x,Q^2)$ spin-dependent structure function as well as
the spin-independent structure functions $F_1(x,Q^2)$
and $F_2(x,Q^2)$, only twist-2 operators contribute
 in the leading order of $1/Q^2$ expansion~\cite{HM}.
On the other hand, not only twist-2 operators but also twist-3 operators
contribute to $g_2(x,Q^2)$ structure function in the leading
order \cite{ARS,KOD1}.

In refs.~\cite{SVETAL,JAFFE}, it is pointed out that
the operators which are proportional to the
{\it equation of motion}~(EOM operator)
appear in the twist-3 operators.
Due to the presence of the EOM operators, the analysis of $g_2(x,Q^2)$
structure function becomes more complicated than the
other structure functions $F_1(x,Q^2)$, $F_2(x,Q^2)$ and $g_1(x,Q^2)$.
The appearance of the EOM operators is a general feature
in the higher-twist operators.

In this talk, we discuss the structure function  $g_2(x,Q^2)$
in the framework of the operator product expansion (OPE)
and renormalization group~\cite{KOD1}.
We focus our attention on the twist-3 operators
which contribute to  $g_2(x,Q^2)$.
Since the twist-3 operators are not all
independent, we need to choose the independent
operator basis to calculate the anomalous
dimensions or the renormalization constants for the twist-3
operators.
We study the operator mixing problem with the EOM
operators being kept. We clarify the role of EOM operators in the course
of the renormalization and point out that the coefficient functions
depend upon the choice of the independent operator basis.

\section{Definition of $g_2(x,Q^2)$}

Let us  consider the polarized  deep inelastic
 lepton-nucleon scattering (fig.1).
%
\input epsf.sty
\begin{center}
\hspace{0.2cm}
\epsfxsize=10cm
\epsfxsize=5cm
\epsffile{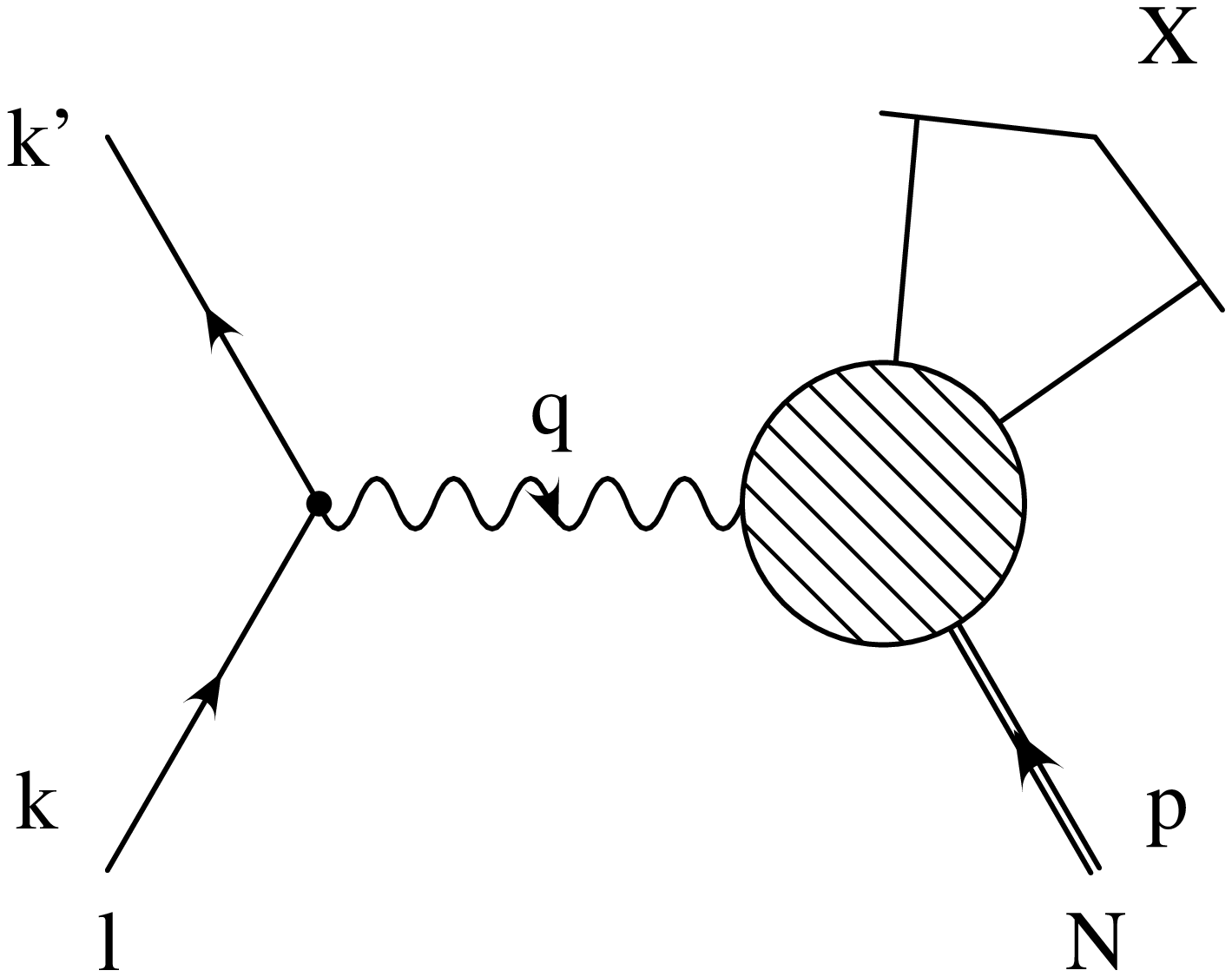}\\
%
{\bf fig.1}~~
$l~+~N~\Longrightarrow ~l~+~X $
\end{center}
%
Here we consider only the electro-magnetic interaction
between the lepton and nucleon.
The kinematical variables $p$, $q$, $k$ and $k'$ are defined in fig.1 and
$Q^2~=~-q^2$. The cross section of this process is given by
\[
k'_0{d\sigma\over d^3k'}~=~{1\over k\cdot p}
 \left({e^2\over 4\pi Q^2}\right )^2 L_{\mu\nu}W^{\mu\nu},
\]
where $L_{\mu\nu}$ and $W_{\mu\nu}$ are the  leptonic and hadronic tenser.
\bea
  L_{\mu\nu}
   &\equiv&{1\over 2}\langle k|j_\mu(0)|k'\rangle
    \langle k'|j_\nu(0)|k\rangle \nonumber\\
  W_{\mu\nu}
   &\equiv& {1\over 2\pi}\sum_X \langle p|J_\mu(0)|X\rangle
    \langle X|J_\nu(0)|p\rangle
      (2\pi)^4\delta^4(p_X-p-q) \nonumber\\
   &=&{1\over 2\pi}\int d^4xe^{iq\cdot x}
     \langle p|[J_\mu(x),J_\nu(0)]|p\rangle.\nonumber
\eea
To define the $g_2$ structure function, we rewrite the
hadronic tensor as follows,
\[
W_{\mu\nu}~\equiv~ W^S_{\mu\nu}~+~iW^A_{\mu\nu},
\]
where $W^S_{\mu\nu}$ is the symmetric part in the Lorentz
indices $\mu\nu$ of the hadronic
tensor which is described by the two spin independent
structure functions $F_1(x,Q^2)$ and $F_2(x,Q^2)$ as,
\[
  W^S_{\mu\nu}=
    - \Big(g_{\mu\nu}-{q^\mu q^\nu \over q^2}\Big)F_1
    +\Big(p_\mu-{p\cdot q\over q^2}q_\mu\Big)
      \Big(p_\nu-{p\cdot q\over q^2}q_\nu\Big)
       {2\over\nu M}F_2~~.
\]
$W^A_{\mu\nu}$ is the antisymmetric part of the hadronic tensor.
We can express the antisymmetric part of the hadronic tensor
with the two spin dependent structure functions
$g_1(x,Q^2)$ and $g_2(x,Q^2)$.
\[
W^A_{\mu\nu}~=~\varepsilon_{\mu\nu\lambda\sigma}q^\lambda
   \left\{ s^\sigma {g_1\over p\cdot q}
    + (p\cdot qs^\sigma - q\cdot s p^\sigma)
      {g_2\over(p\cdot q)^2}\right\}.
\]
$x$ is the Bjorken variable
given by $x=Q^2/2p\cdot q=Q^2/2M\nu$ ,
$p\cdot q=M\nu$. $M$ is the nucleon mass and
$s$ is the covariant spin vector defined by
$s^\mu~=~\overline{u}(p,s)\gamma^\mu\gamma_5 u(p,s)$.

\section{Operator product expansion}

To apply the OPE to the polarized deep inelastic
leptoproduction, we introduce the corresponding
forward virtual Compton amplitude $T_{\mu\nu}$ \cite{KODT}.
\[
T_{\mu\nu}~=~i\int dx e^{iq\cdot x}
\langle p,s|TJ_\mu(x)J_\nu(0)|p,s\rangle.
\]
{}From the optical theorem, $T_{\mu\nu}$ and $W_{\mu\nu}$ are
related through the relation,
\[
  W_{\mu\nu}~=~{1\over\pi}
    {\rm Im} T_{\mu\nu}.
\]

The antisymmetric part of the currents
product can be expanded with composite operators and
Wilson's coefficient functions as,
\bea
i\int d^4xe^{iq\cdot x} T(J_{\mu}(x)J_{\nu}(0))^A
    &=& -i\varepsilon_{\mu\nu\lambda\sigma} q^{\lambda}
        \sum_{n=1,3,\cdots} \left( \frac{2}{Q^2 }\right )^n
           q_{\mu _1} \cdots q_{\mu _{n-1}} \nonumber\\
    & &  \qquad \times \Bigl \{ E_q^n R_q^{\sigma\mu_{1}\cdots \mu_{n-1}}
        + \sum_j E_j^n R_j^{\sigma\mu_{1}\cdots \mu_{n-1}} \Bigr \} ,
\label{op2}
\eea
where $R_i^n$'s are the composite operators and $E_i^n$'s are
the corresponding coefficient functions.
In (\ref{op2}), $R_q$ represents the twist-2 operators
and the other operators inside the summation over $j$
are the twist-3 operators. For simplicity, let us consider
the flavor non-singlet case. (In the following, we omit the flavor matrices
for the quark fields.)
The explicit form of the twist-2 operator is given by
\[
  R_q^{\sigma\mu_{1}\cdots \mu_{n-1}} =
         i^{n-1} \overline{\psi}\gamma_5 \gamma^{\{\sigma}D^{\mu_1}
              \cdots D^{\mu_{n-1}\}}\psi \ -{\rm (traces)} \ ,
\]
where $\{ \quad \}$ means symmetrization over the Lorentz indices
and $-$ (traces) stands for the subtraction of the trace terms to
make the operators traceless. (The trace terms will be omitted
in the following.)
For the twist-3 operators, we have
\bea
  R_F^{\sigma\mu_{1}\cdots \mu_{n-1}} &=&
         \frac{i^{n-1}}{n} \Bigl[ (n-1) \overline{\psi}\gamma_5
       \gamma^{\sigma}D^{\{\mu_1} \cdots D^{\mu_{n-1}\}}\psi
\nonumber\\
   & & \qquad\qquad - \sum_{l=1}^{n-1} \overline{\psi} \gamma_5
       \gamma^{\mu_l }D^{\{\sigma} D^{\mu_1} \cdots D^{\mu_{l-1}}
            D^{\mu_{l+1}} \cdots D^{\mu_{n-1}\}}
                             \psi \Bigr] ,
                                            \label{op3}\\
  R_m^{\sigma\mu_{1}\cdots \mu_{n-1}} &=&
          i^{n-2} m \overline{\psi}\gamma_5
       \gamma^{\sigma}D^{\{\mu_1} \cdots D^{\mu_{n-2}}
        \gamma ^{\mu_{n-1}\}} \psi ,
                                             \label{op4} \\
  R_k^{\sigma\mu_{1}\cdots \mu_{n-1}} &=& \frac{1}{2n}
              \left( V_k - V_{n-1-k} + U_k + U_{n-1-k} \right) ,
                                                  \label{op5}
\eea
where $m$ in (\ref{op4}) represents the quark mass (matrix).
The operators in (\ref{op5}) contain
the gluon field strength $G_{\mu\nu}$ and it's dual tensor
$\widetilde{G}_{\mu \nu}={1\over
2}\varepsilon_{\mu\nu\alpha\beta}
G^{\alpha\beta}$ explicitly and are given by
\bea
    V_k &=& i^n g S \overline{\psi}\gamma_5
       D^{\mu_1} \cdots G^{\sigma \mu_k } \cdots D^{\mu_{n-2}}
        \gamma ^{\mu_{n-1}} \psi , \nonumber \\
    U_k &=& i^{n-3} g S \overline{\psi}
       D^{\mu_1} \cdots \widetilde{G}^{\sigma \mu_k } \cdots
             D^{\mu_{n-2}} \gamma ^{\mu_{n-1}} \psi , \nonumber
\eea
where $S$ means symmetrization over $\mu_i$ and $g$ is the QCD
coupling constant.

It is well-known that these operators (\ref{op3})-(\ref{op5}) are not
independent and related through  EOM operator .
\bea
   R_{eq}^{\sigma\mu_{1}\cdots \mu_{n-1}}
    &=&   i^{n-2} \frac{n-1}{2n} S [ \overline{\psi} \gamma_5
             \gamma^{\sigma} D^{\mu_1} \cdots D^{\mu_{n-2}}
               \gamma ^{\mu_{n-1}} (i\not{\!\!D} - m )\psi \nonumber\\
    & & \qquad\qquad\qquad\qquad + \overline{\psi} (i\not{\!\!D} - m )
             \gamma_5 \gamma^{\sigma} D^{\mu_1} \cdots D^{\mu_{n-2}}
                 \gamma ^{\mu_{n-1}} \psi ] \nonumber  .
\eea
Making use of the identities $D_\mu={1\over2}\{\gamma_\mu,\not{\!\!\!D}\}$
and $[D_\mu,D_\nu]=g G_{\mu\nu}$,
we can obtain the following relation for the twist-3 operators,
\be
     R_F^{\sigma\mu_{1}\cdots \mu_{n-1}} =
       \frac{n-1}{n} R_m^{\sigma\mu_{1}\cdots \mu_{n-1}}
         + \sum_{k=1}^{n-2} (n-1-k)
            R_k^{\sigma\mu_{1}\cdots \mu_{n-1}} +
              R_{eq}^{\sigma\mu_{1}\cdots \mu_{n-1}} .
\label{op6}
\ee

\section{Operator mixing problem}
Now we investigate the $Q^2$ evolution by using the
renormalization group method.
The presence of the EOM operators brings about some complication
through the course of renormalization.

Although the physical matrix elements of the EOM operators vanish
{}~\cite{POLI,COLL},
we must keep them to study the renormalization
because their off-shell Green's function do not vanish.
We analyze the operator mixing problem by keeping the EOM operators.
In ref.\cite{KU}
we examined what happens to the renormalization if there are several
operators and these operators are related by constraints.

Here we show the one-loop results for the $n=3$ case  as the simplest example.
In this case we have four operators with the constraint,
\be
  R_F = \textstyle{\frac{2}{3}} R_m + R_1 + R_{eq},
\label{n3}
\ee
where the Lorentz indices of operators are omitted.

First we choose the operators $R_F, R_m, {\rm and}R_1$ as independent operators
and eliminate EOM operator $R_{eq}$.
We get the following renormalization matrix for the composite operators,
\be
\bmat{c}
R_F \\ R_1 \\ R_m \\ R_{eq1}
\emat_{\hspace{-0.1cm}R}
=
\bmat{cccc}
Z_{11} & Z_{12} & Z_{13} & Z_{14} \\
Z_{21} & Z_{22} & Z_{23} & Z_{24} \\
0 & 0 & Z_{33} & 0 \\
0 & 0 & 0 & Z_{44}
\emat
\bmat{c}
R_F \\ R_1 \\ R_m \\ R_{eq1}
\emat_{\hspace{-0.1cm}B}
\label{n3z1}
\ee
where
$Z_{ij}$ are given in the
dimensional regularization
$D=4-2\varepsilon$:
\[
Z_{ij}\equiv \delta_{ij} +
{1\over\varepsilon}{{g^2}\over{16\pi^2}}z_{ij}
\]
A straightforward but tedious calculation gives as,
\be
\ba{ll}
z_{11}={7\over 6}C_2(R)+{3\over 8}C_2(G), &
z_{12}=-{3\over 2}C_2(R)+{21\over 8}C_2(G), \\
&\\
z_{13}=3C_2(R)-{1\over 4}C_2(G), & z_{14}=-{3\over 8}C_2(G), \\
&\\
z_{21}={1\over 6}C_2(R)-{1\over 8}C_2(G), &
z_{22}=-{1\over 2}C_2(R)+{25\over 8}C_2(G), \\
&\\
z_{23}=-{1\over 3}C_2(R)+{1\over 12}C_2(G), &
z_{24}={1\over 8}C_2(G), \\
&\\
z_{33}=6C_2(R), & z_{44}=0 .
\label{zn3}
\ea
\ee
The quadratic Casimir operators are $C_2(R)=4/3$ and $C_2(G)=3$
for the case of QCD.
$R_{eq1}$ is a gauge non-invariant EOM operator.
\[
  R_{eq1}^{\sigma\mu_{1} \mu_{2}} =i \textstyle{1\over 3} S
  [ \overline{\psi} \gamma_5 \gamma^{\sigma} \partial ^{\mu_1}
  \gamma ^{\mu_{2}} (i\not{\!\!D} - m )\psi + \overline{\psi}
  (i\not{\!\!D} - m ) \gamma_5 \gamma^{\sigma} \partial ^{\mu_1}
  \gamma ^{\mu_{2}} \psi ] .
\]
Although this operator is gauge non-invariant,
it is possible to appear in the operator basis
because it vanishes by the equation of motion~\cite{COLL,KU}.
This result satisfies the equalities.
\be
\ba{ll}
z_{11}+z_{12} = z_{21}+z_{22}, &
    {2\over 3}z_{11}+z_{13} = {2\over 3}z_{21}+z_{23}+{2\over 3}z_{33}, \\
&\\
  z_{13}-{2\over 3}z_{12} = z_{23}-{2\over 3}z_{22}+{2\over 3}z_{33}.&
\ea
\label{zrel}
\ee

What happens if we choose $R_1$, $R_m$, $R_{eq}$ and $R_{eq1}$,
and eliminate $R_F$?
This choice is the same as one adopted by the authors in ref.\cite{SVETAL}.
Using (\ref{n3}) and relations (\ref{zrel})
we get the renormalization matrix
\be
\bmat{c}
R_1 \\ R_m \\ R_{eq} \\ R_{eq1}
\emat_{\hspace{-0.1cm}R}
=
\bmat{cccc}
Z_{21}+Z_{22} & {2\over 3}Z_{21}+Z_{23} & Z_{21} & Z_{24} \\
0 & Z_{33} & 0 & 0 \\
0 & 0 & Z_{11}-Z_{21} & Z_{14}-Z_{24} \\
0 & 0 & 0 & Z_{44}
\emat
\bmat{c}
R_1 \\ R_m \\ R_{eq} \\ R_{eq1}
\emat_{\hspace{-0.1cm}B} ,
\ee
where
$Z_{ij}$ are defined in (\ref{n3z1}) and (\ref{zn3}). In this basis, our
results for $n=3$ case agree with those in ref.\cite{SVETAL}.

It is to be noted that the renormalization matrix for the operators
including EOM operators becomes triangular
because the counter-term to the operator $R_{eq}$ should vanish by the
equation of motion\cite{COLL}.
Our results are consistent with this general argument.

\section{Coefficient function}

Next we determine the coefficient functions at the tree level.
We used the technique, discussed by E.V. Shuryak and A.I. Vainshtein
\cite{SVETAL} and R. L. Jaffe and M. Soldate \cite{SOLD}.
The coefficient function can be obtained by the short distance expansion
of the current products in the presence of the external gauge field.

We include the fermion bilinear operators $R_F$ in the operator basis.
We have the coefficient functions,
\be
   E_q^n(tree)~=~E_F^n(tree)~=~1,  E_m^n(tree)~=~E_k^n(tree)~=~0.
\ee
On the other hand, if we eliminate $R_F$ in the basis,
and we use the constraints of the twist-3 operators (\ref{n3}), we have,
\be
  E_q^n ({\rm tree})=1\,,\quad E_m^n ({\rm tree})= {{n-1}\over n}\,,
   \quad E_k^n ({\rm tree}) = n-1-k.
\label{coef2}
\ee

Now let us see the moment sum rules.
General form of the moment for $g_2$ spin structure function is given by,
\bea
  M_n \equiv \int_0^1 dx x^{n-1} g_2 (x,Q^2)
    &=& - {{n-1}\over {2n}} \Bigl [ a_n E_q^n(Q^2)
         - d_n E_F^n(Q^2) \Bigr ] \nonumber\\
    & & \qquad\qquad + {1\over 2} \Bigl [ e_n E_m^n(Q^2)
         + \sum_{k=1}^{n-2}f_n^k E_k^n(Q^2) \Bigr ].
\label{g2sumrule}
\eea
Here $a_n$,$d_n$,$e_n$ and $f^k_n$ are matrix
elements of the operators sandwiched between nucleon states with
momentum $p$ and spin $s$,
\bea
  \langle p,s | R_q^{\sigma\mu_{1}\cdots \mu_{n-1}} |p,s \rangle
       &=& - a_n s^{\{\sigma}p^{\mu_1} \cdots p^{\mu_{n-1}\}}
                           \label{element1}\\
  \langle p,s | R_F^{\sigma\mu_{1}\cdots \mu_{n-1}} |p,s \rangle
       &=& -  \frac{n-1}{n} d_n ( s^{\sigma}p^{\mu_1} - s^{\mu_1}p^{\sigma})
                    p^{\mu_2} \cdots p^{\mu_{n-1}} \\
  \langle p,s | R_m^{\sigma\mu_{1}\cdots \mu_{n-1}} |p,s \rangle
       &=& - e_n ( s^{\sigma}p^{\mu_1} - s^{\mu_1}p^{\sigma})
                    p^{\mu_2} \cdots p^{\mu_{n-1}} \\
  \langle p,s | R_k^{\sigma\mu_{1}\cdots \mu_{n-1}} |p,s \rangle
       &=& - f_n^k ( s^{\sigma}p^{\mu_1} - s^{\mu_1}p^{\sigma})
                    p^{\mu_2} \cdots p^{\mu_{n-1}} \\
  \langle p,s | R_{eq}^{\sigma\mu_{1}\cdots \mu_{n-1}} |p,s \rangle
      &=& 0 \label{elementeq}.
\eea
{}From the constraint for the twist-3 operators (\ref{op6}), we get relation,
\be
    {{n-1}\over n}d_n={{n-1}\over n}e_n+\sum_{k=1}^{n-2}(n-1-k)f_n^k.
\label{co1}
\ee

If we choose $R_k$ and $R_m$ as the independent operator basis
and use the relation (\ref{co1}), we have the moments for $g_2$
with the coefficient functions given in(\ref{coef2}),
\be
    M_n = - {{n-1}\over {2n}} a_n E_q^n(Q^2)
    + {1\over 2} \Bigl [ e_n E_m^n(Q^2)
         + \sum_{k=1}^{n-2}f_n^kE_k^n(Q^2) \Bigr ].
\label{moment}
\ee

We next check this result by looking at the explicit form of the both
side at order of $g^2$.
Using the perturbative solution of the renormalization group
equation,
the right-hand side of (\ref{moment}) becomes for the quark matrix elements
in the leading order of $\ln Q^2$,
\be
{\rm RHS \ of \ }(\ref{moment}) = - {1\over 2}{g^2\over{16\pi^2}}
      \left ( {{n-1}\over n}\Bigl( -{1\over 2}\Bigr) \gamma_q^0
            E_q^n ({\rm tree})+{1\over 2}\gamma_{mk}^0
                 E_k({\rm tree}) \right ) \ln{Q^2} + \cdots .
\label{g2ope}
\ee
On the other hand, we can get the $\ln Q^2$ dependence
by calculating the one-loop Compton amplitude off the on-shell
massive quark target.
In the leading order of  $\ln Q^2$,
the moments becomes,
\be
    M_n = {1\over 2}{g^2\over{16\pi^2}}C_2(R)(-2+{4\over{n+1}})\ln{Q^2}
                   +\cdots .
\label{comp}
\ee
{}From the off-diagonal element of the renormalization constant
matrix $Z_{km}$, the anomalous dimension $\gamma_{mk}$ reads,
\be
\gamma_{mk}\equiv -{{g^2}\over {16\pi^2}}{{8C_2(R)}\over n}
{1\over{k(k+1)(k+2)}}\equiv {{g^2}\over {16\pi^2}}\gamma_{mk}^0.
\label{gammk}
\ee
This result is in disagreement with the one given in the fifth
reference in \cite{SVETAL}.
The expression for the anomalous dimensions for
the twist-2 operators $R_F$ and the quark mass dependent
twist-3 operators $R_m$ are given by,
\be
\gamma_q^0 = 2C_2(R) \Bigl [ 1-{2\over{n(n+1)}} +
   4\sum_{k=2}^n{1\over k} \Bigr ],\qquad
          \gamma_{mm}^0=8C_2(R) \sum_{k=1}^{n-1}{1\over k}.
\label{gammm}
\ee
Putting (\ref{gammk}) and (\ref{gammm})
into the above equation
with the tree level coefficient functions (\ref{coef2}), we find
(\ref{g2ope}) coincides with (\ref{comp}).
Thus we confirm that our results(\ref{coef2}) and (\ref{gammk})
are consistent with the explicit calculation.

\section{Summary}

We investigate the $g_2(x,Q^2)$ spin structure function in
the framework of the OPE. For $g_2(x,Q^2)$, the twist-3
operators contribute in the leading order of the $1/Q^2$ expansion.
The twist-3 operators are not independent but constrained through
the EOM operators which vanish on application of the
equation of motion. Here we demonstrate an analysis for
the mixing of the twist-3 operators by keeping the EOM operators.
We also determine the coefficient function at the tree level.
We pointed out that the coefficient functions depend upon the choice of the
independent operator basis.

\noindent{\large{\bf Acknowledgment}}\\
\thanks{
This work is being undertaken with
Prof. T. Uematsu and Prof. J. Kodaira .
The author would like to thank his collaborators and
Prof. K. Sasaki for discussions.
The author also wishes to thank Prof. S. Parke
and Dr. T. Onogi for the useful comments.
This work is supported in part by the Monbusho
Grant-in-Aid for Scientific Research No. 050076.}



\end{document}